\documentclass[aps,prl,twocolumn,amsmath,amssymb,superscriptaddress]{revtex4}
\usepackage{graphicx}
\usepackage{Jonasmacros}
\usepackage{mathptmx}
\usepackage{txfonts}
\usepackage{bm,bbold} 
\usepackage[colorlinks,citecolor=cyan,linkcolor=Fuchsia]{hyperref}
\usepackage[usenames,dvipsnames,svgnames]{xcolor}

\newcommand{\tobedeleted}[1]{\textcolor{green}{#1}}
\renewcommand{\tobedeleted}[1]{\relax}

\begin{document}
\date{\today}
\title{\textcolor{black}{Remote Mesoscopic Signatures of Induced Magnetic Texture in Graphene}}

\author{N. Arabchigavkani}
\email{nargessa@buffalo.edu}
\affiliation{Department of Physics, University at Buffalo, the State University of New York, Buffalo, New York 14260, USA}
\author{R. Somphonsane}
\author{H. Ramamoorthy}
\affiliation{Department of Physics, King Mongkut's Institute of Technology Ladkrabang, Bangkok, 10520, Thailand}
\author{G. He}
\author{J. Nathawat}
\author{S. Yin}
\affiliation{Department of Electrical Engineering, University at Buffalo, the State University of New York, Buffalo, New York 14260, USA}
\author{B. Barut}
\affiliation{Department of Physics, University at Buffalo, the State University of New York, Buffalo, New York 14260, USA}
\author{K. He}
\author{M. D. Randle}
\author{R. Dixit}
\affiliation{Department of Electrical Engineering, University at Buffalo, the State University of New York, Buffalo, New York 14260, USA}
\author{K. Sakanashi}
\author{N. Aoki}
\affiliation{Department of Materials Science, Chiba University, Chiba, 263-8522, Japan}
\author{K. Zhang}
\author{L. Wang}
\affiliation{Department of Materials Science and Engineering, University of Science and Technology of China, Hefei, PR China}
\author{W.-N. Mei}
\affiliation{Department of Physics, University of Nebraska Omaha, Omaha, NE 68182, USA}
\author{P. A. Dowben}
\affiliation{Department of Physics and Astronomy, Theodore Jorgensen Hall, University of Nebraska Lincoln, Lincoln, Nebraska 68588-0299, USA}
\author{J. Fransson}
\affiliation{Department of Physics and Astronomy, Uppsala University, Box 534, SE-751 21, Uppsala, Sweden}
\author{J. P. Bird}
\email{jbird@buffalo.edu}
\affiliation{Department of Electrical Engineering, University at Buffalo, the State University of New York, Buffalo, New York 14260, USA}

\date{\today}

\begin{abstract}
Mesoscopic conductance fluctuations are a \textcolor{black}{ubiquitous} signature of phase-coherent transport in small conductors, exhibiting universal character independent of system details. In this work, however, we demonstrate a pronounced breakdown of this universality, due to the interplay of local and \textcolor{black}{remote} phenomena in transport. Our experiments are performed in \textcolor{black}{a graphene-based \emph{interaction-detection}} geometry, in which an \textcolor{black}{artificial magnetic texture is induced in the graphene layer} by covering a portion of it with a micromagnet. \textcolor{black}{When probing conduction at some distance from this region, the strong influence of remote factors is manifested through the appearance of giant conductance fluctuations, with amplitude much larger than $e^2/h$. This violation of one of the fundamental tenets of mesoscopic physics dramatically demonstrates how local considerations can be overwhelmed by remote signatures in phase-coherent conductors.}
\end{abstract}
\maketitle

As the size of conductors is reduced towards the fundamental scales governing electron transport, their electrical behavior is dramatically modified \cite{Ferry2009}. In this mesoscopic regime, the wavelike nature of carriers causes Drude conduction to be overwhelmed by quantum-interference phenomena, most notable of which are weak localization \cite{Bergmann1984} and universal conductance fluctuations \cite{Lee1987,Washburn1986} (UCF). Guided by the Landauer formalism, a quantitative understanding of these effects was achieved many decades ago. Most notably, the UCF are a signature of interference among the different Feynman paths for transmission and exhibit a maximum amplitude of $e^2/h$, independent of system size or the degree of disorder \cite{Lee1987}. This universality has been confirmed in experiments on both metals and semiconductors, long providing the perspective from which our understanding of mesoscopic transport is derived.

Another consequence of phase coherence in mesoscopic transport is \textcolor{black}{the breakdown of classical locality, allowing the conductance of small systems to be influenced by \emph{remote} processes, arising outside of the region under study. Examples include non-local voltage fluctuations in metal wires} \cite{Washburn1986}, Fano resonances in systems of coupled quantum point contacts \cite{Yoon2012,Fransson2014}, and long-range flavor currents in graphene \cite{Abanin2011}.

In this Letter, \textcolor{black}{we address the question of what happens when phase-coherent transport in a mesoscopic system is determined by an \emph{interplay} of local and remote considerations, and demonstrate how coexistence of these effects can lead to conductance features that violate established universality. To realize this behavior, we perform experiments on a conducting graphene sheet that is partially covered by a ferromagnet (see Fig. 1(a)). In the \emph{interaction region} in which the graphene directly contacts the magnet, carriers in the atomically thin carbon layer experience strong magnetic interactions, as the bandstructures of the two materials hybridize \cite{Karpan2007,Varykhalov2008,Marchenko2012,Calleja2014,Marchenko2015,Usachov2015,Yang2016}. Net spin polarization is consequently induced in the carbon sheet \cite{Marchenko2015,Usachov2015}, and should be preserved (due to weak spin decoherence \cite{Yang2011,Han2011,Dlubak2012,Han2014}) when carriers drift away from this region, towards an uncovered section of native graphene (the \emph{detection region}). To explore this scenario, we perform measurements of the differential conductance of the detection region, and reveal the presence of features that definitively arise from the influence of the magnetic element. This influence is manifested as complex structure in the differential conductance, both at and around zero bias, that is absent when the interaction region is excluded from the circuit. As the carrier concentration in the graphene is varied, this remote mesoscopic signal generates giant fluctuations in the conductance, with an amplitude that exceeds the normal universal value \cite{Lee1987} by well over an order of magnitude. This violation of one of the fundamental tenets of mesoscopics dramatically demonstrates how local considerations can be overwhelmed by remote signatures in phase-coherent conductors.}


\begin{figure}[t]
\begin{center}
\includegraphics[width=0.99\columnwidth]{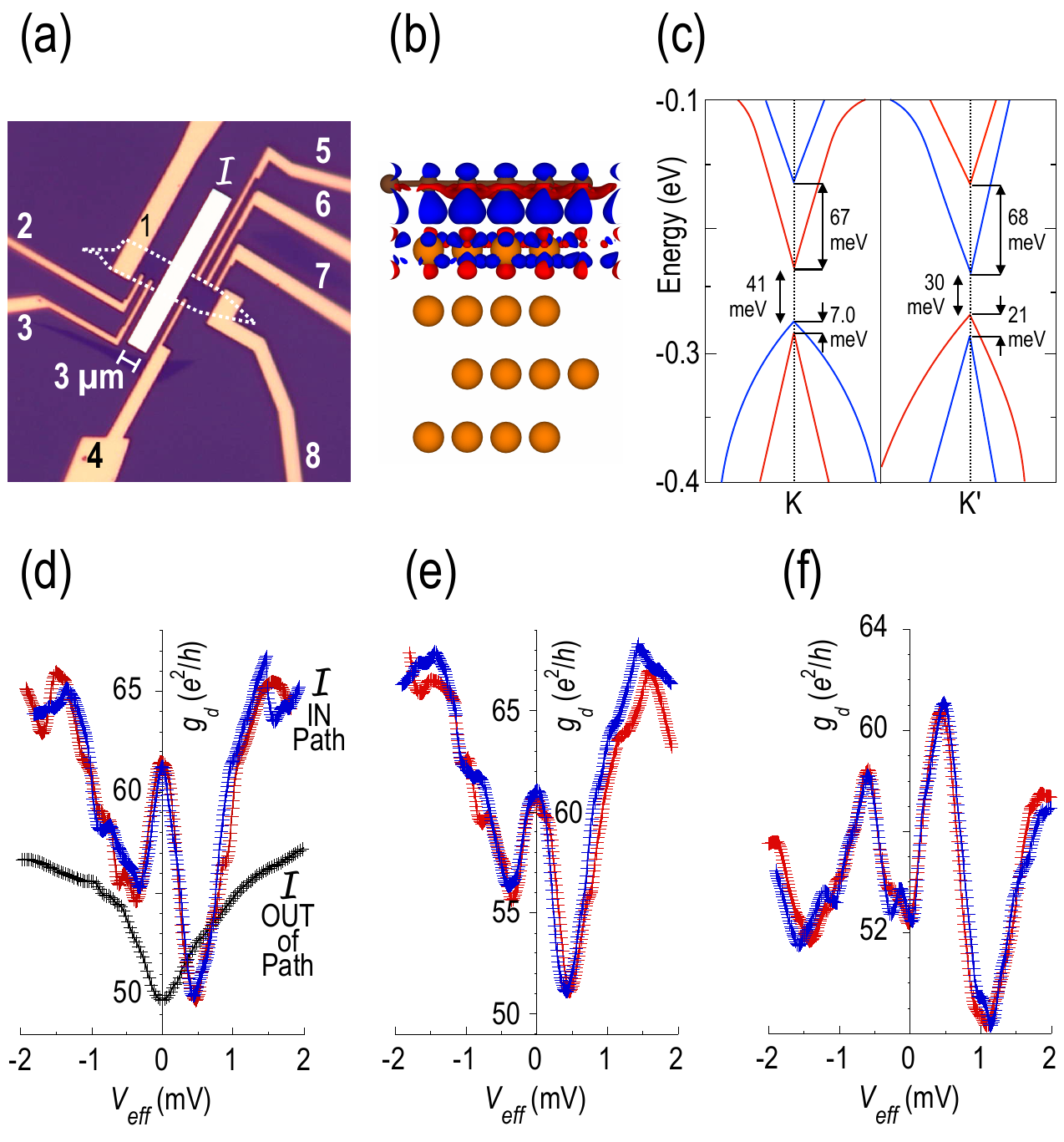}
\end{center}
\caption{
(Color online) (a) Optical micrograph showing the \textcolor{black}{interaction-detection geometry of device} I-D:1. Electrodes 1 $-$ 8 are ohmic (Cr/Au: 5-/75-nm) contacts, used to make the differential-conductance measurements. \textcolor{black}{The (20-nm thick) Co ferromagnet that defines the interaction region is denoted as $\mathcal{I}$}, and the shape of the graphene sheet is enclosed by the white dotted line. (b) Computed charge density difference at the graphene/Co(111) interface. Electron loss (enrichment) is displayed in red (blue). The graphene layer (at top) is indicated by brown spheres and the yellow spheres represent Co atoms. (c) Calculated bandstructure of graphene near the K and K$^\prime$ points. Red (blue) bands correspond to spin-down (spin-up) states. (d) Differential conductance measured in the interaction-detection geometry, comparing the effect of including (red/blue data) or excluding (black data) the interaction region ($\mathcal{I}$) in the measurement path. Panels (e) \& (f) illustrate the manner in which the structure around zero bias changes as the gate voltage is varied. All data in panels (d) $-$ (f) are from I-D:1, with red (blue) data obtained while sweeping the bias voltage up (down). (d) $V_{g} - V_{D}$ = $-$114 V. (e) $V_{g} - V_{D}$ = $-$113 V. (f) $V_{g} - V_{D}$ = $-$100 V.
}
\label{Fig1}
\end{figure}

An example of \textcolor{black}{the interaction-detection geometry} is provided in Fig. 1(a), where a graphene sheet is partially covered by a floating Co element (\textcolor{black}{$\mathcal{I}$}) that defines the interaction region. \textcolor{black}{When graphene is placed in intimate contact with a ferromagnet, hybridization of the electronic structure of the two materials occurs \cite{Karpan2007,Varykhalov2008,Marchenko2012,Calleja2014,Marchenko2015,Usachov2015,Yang2016}. This has profound implications for the carriers in graphene, which can spin polarize while also experiencing an external spin-orbit coupling. Some of these features are highlighted in Figs. 1(b) \& 1(c), in which we show electronic structure calculations \textcolor{black}{performed for graphene-on-Co(111)} (see Section S6.1 of the Supplementary Material for details).}
Figure 1(b) shows strong hybridization of the $p_z$ orbitals of graphene with Co, an interaction that modifies the bandstructure of the graphene as shown in Fig. 1(c). To understand this bandstructure, which is plotted in Section S6.1 of the supplement for a wider range of energy, and which \textcolor{black}{exhibits similarities with the results of prior studies of graphene-on-Co(1000)} \cite{Marchenko2015,Usachov2015}, in Section S6.2 we present an effective model for the hybrid graphene system. This includes terms in the Hamiltonian describing spin-orbit coupling and exchange induced spin-polarization, allowing us to separately identify their contributions. From this we conclude that the gaps (of 30 $-$ 40 meV) and spin-splitting in the bandstructure of Fig. 1(c) result from the combined influence of spin-orbit interaction and exchange coupling. A crucial point that should also be noted is that the usual degeneracy of the K and K$^\prime$ points in graphene is lifted in the presence of the ferromagnet, which breaks inversion symmetry. The bands near these points are therefore inequivalent and lead, consequently, to non-zero carrier spin polarization; this is calculated explicitly in the supplement (see Figs. S9 \& S11), where we predict that it attains a value of $>$80\% over a wide range of energy around the Fermi level.


Basic details of our devices are described in Sections S1 \& S3.1 of the supplement. While we focus here on results obtained from a systematic study of device \textcolor{black}{I-D:1}, our observations have been confirmed in a second structure (\textcolor{black}{I-D:2}, see Section S5). Four-probe differential-conductance ($g_d$) was measured by superimposing a small AC voltage upon a larger DC component ($V_d$), and in our analysis we plot the variation of $g_d$ as a function of the portion of that voltage ($V_\text{\emph {eff}}$) dropped \cite{Somphonsane2020} across the graphene itself
\textcolor{black}{(see Section S2 of the Supplementary Material)}.
Variation of the voltage ($V_g$) on the Si substrate was used to tune the carrier density. All measurements were made with the devices mounted in the vacuum chamber of a closed-cycle cryostat that, unless stated otherwise, was operated at a stable base temperature of 3 K. \textcolor{black}{The measurements were also made at zero external magnetic field, where shape anisotropy favors an in-plane multi-domain state close to that of a bar magnet (see Section S3.2 of the supplement).}

The key aspects of our study are highlighted in Fig. 1(d), which shows the differential conductance of the \emph{same} section of graphene, with and without the magnetic element included in the current path. Black data correspond to the latter case, where the external (AC \& DC) voltages are applied across probes 5 \& 8 in Fig. 1(a) and the differential conductance is determined from the voltage across probes 6 \& 7. In this configuration, $g_d$ starts from a local minimum at zero bias, before increasing monotonically when a DC bias of either polarity is applied. Previously, we \textcolor{black}{have used a combination of magneto-transport and differential conductance studies to establish} that this general signature arises from a bias-induced dephasing of weak localization \cite{Somphonsane2020}. Very different behavior occurs, however, with the magnet included in the current path. In this case, we again determine the differential conductance using voltage probes 6 \& 7, but now apply the external (AC \& DC) voltages across probes 1 \& 8 to include $\mathcal{I}$ in the current path (albeit several microns away from the section of graphene we are probing). The differential conductance is dramatically transformed in this geometry, exhibiting a pronounced \emph{peak} near zero bias and mesoscopic fluctuations over a wider range of voltage. The reproducibility of these features is confirmed by the close overlap of red and blue data points in the figure, which correspond to experiments performed while sweeping the bias voltage in opposite directions \cite{Reproducibility}.

\begin{figure}[t]
\begin{center}
\includegraphics[width=0.99\columnwidth]{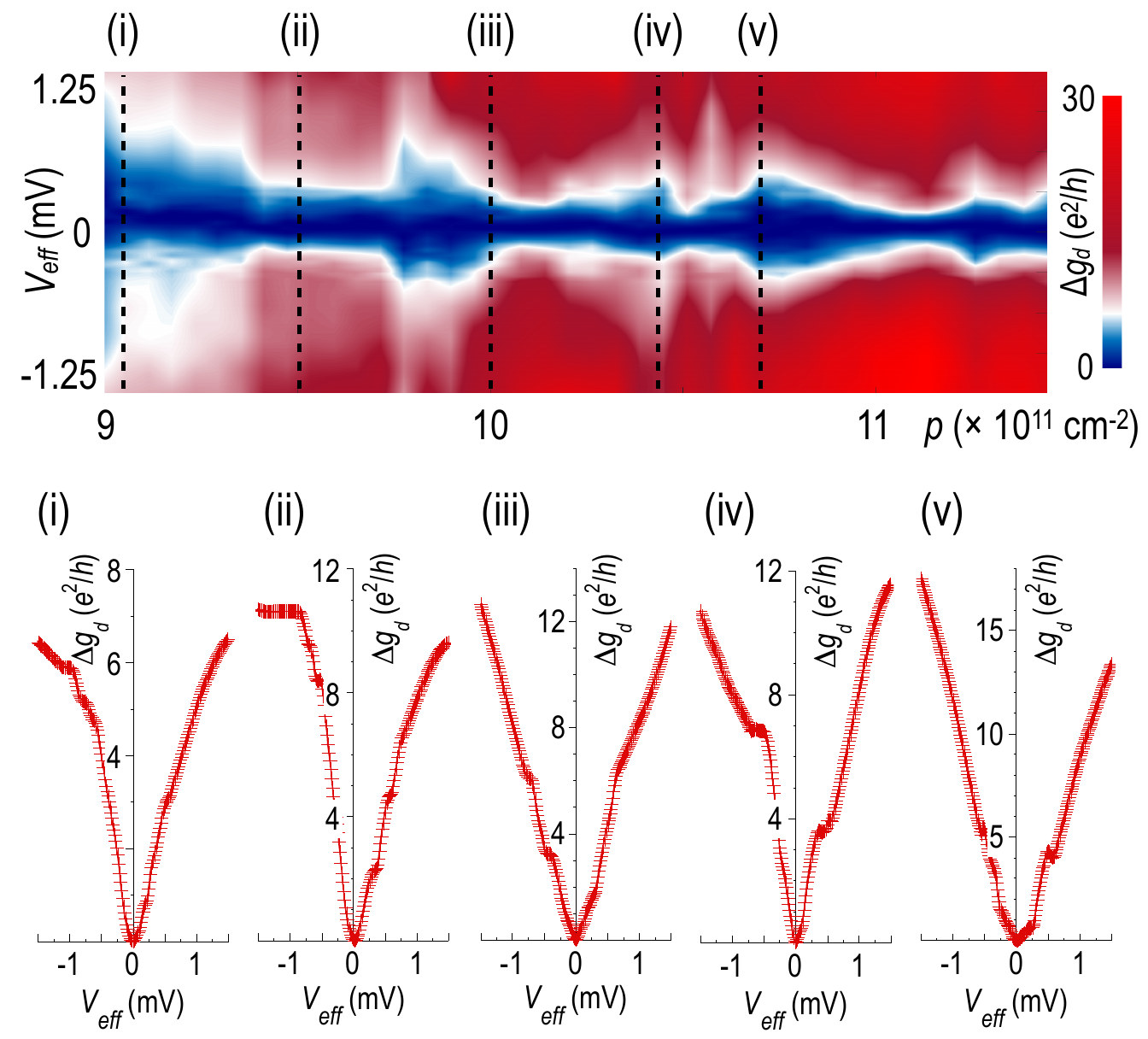}
\end{center}
\caption{
(Color online) Gate-voltage dependent evolution of differential conductance for I-D:1, measured with $\mathcal{I}$ excluded from the current path (external voltages: probes 5 \& 8; $V_\text{\emph {eff}}$: probes 6 \& 7). The lower panels labeled (i) $-$ (v) provide representative examples of the differential conductance at different gate voltages. \textcolor{black}{$\Delta g_{d}$ is defined as the variation of differential conductance, relative to its minimum value ($g_{d_{min}}$, in units of $e^{2}/h$) over the indicated bias range.} (i) $g_{d_{min}}$ = 49.7. (ii) $g_{d_{min}}$ = 50.4. (iii) $g_{d_{min}}$ =54.6. (iv) $g_{d_{min}}$ = 62.3. (v) $g_{d_{min}}$ = 61.5.
}
\label{Fig1}
\end{figure}

With the magnet in the current path, the differential conductance exhibits a complex evolution with carrier concentration. This is apparent from Figs. 1(d) $-$ 1(f), which show measurements performed at different gate voltages. While a zero-bias peak is present in Figs. 1(d) \& 1(e), in Fig. 1(f) this is transformed into a doublet-like structure that is centered around zero-bias. This dramatic change in differential conductance is highly suggestive of the role of quantum fluctuations due to mesoscopic interference, a point that we further demonstrate in Figs. 2 \& 3. Here we plot the variation of differential conductance as a systematic function of hole concentration ($p$), with the magnetic element both excluded from (Fig. 2), and included in (Fig. 3), the current path. Prominent in Fig. 2 is a suppression of conductance around zero bias; this is apparent, also, in the line plots of panels (i) $-$ (v) and, as mentioned earlier, has been identified as a signature of weak localization \cite{Somphonsane2020}. \textcolor{black}{The localization effect yields a conductance minimum at zero-bias for \emph{all} carrier concentrations. The lineshape of the differential conductance (its bias-dependent width and amplitude) does evolve irregularly as the carrier density is varied, but this kind of stochastic response is known from the study of weak localization in other mesoscopic systems \cite{Chang1994,Chan1995,Huibers1998}. The complex variation reflects the fact that, in mesoscopic systems, the localization correction is not self-averaging, but instead varies stochastically with carrier density. Of most importance for the discussion here, the stochastic nature of the contour of Fig. 2 provides a strong indication that transport in the graphene is strongly phase coherent.}

\begin{figure}[t]
\begin{center}
\includegraphics[width=0.99\columnwidth]{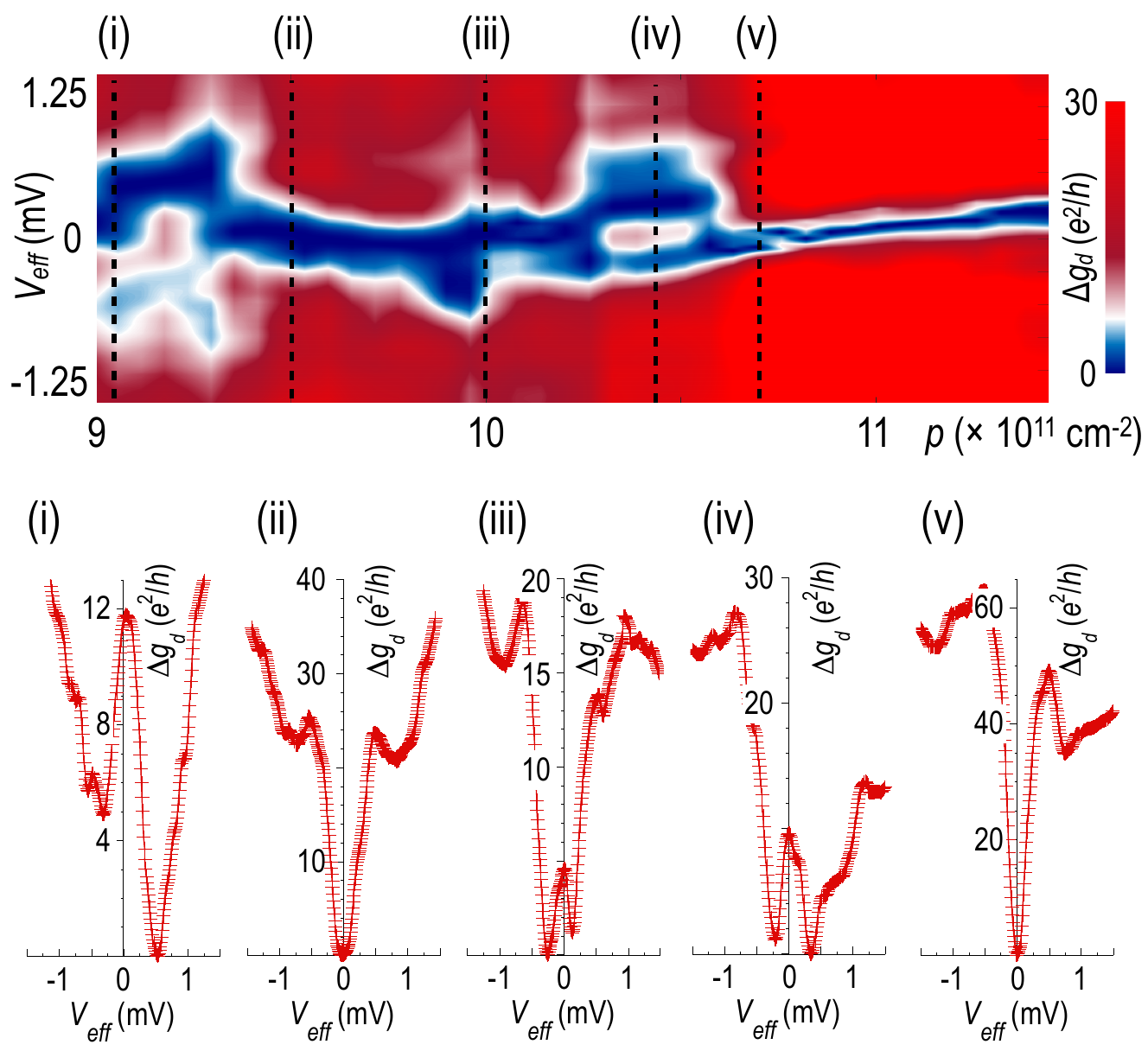}
\end{center}
\caption{
(Color online) Gate-voltage dependent evolution of differential conductance for I-D:1, measured with $\mathcal{I}$ included in the current path (external voltages: probes 1 \& 8; $V_\text{\emph {eff}}$: probes 6 \& 7). The lower panels labeled (i) $-$ (v) provide representative examples of the differential conductance at different gate voltages. \textcolor{black}{$\Delta g_{d}$ is defined as the variation of differential conductance, relative to its minimum value ($g_{d_{min}}$, in units of $e^{2}/h$) over the indicated bias range.} (i) $g_{d_{min}}$ = 52.0. (ii) $g_{d_{min}}$ = 44.8. (iii) $g_{d_{min}}$ = 60.0. (iv) $g_{d_{min}}$ = 66.0. (v) $g_{d_{min}}$ = 40.0.
}
\label{Fig1}
\end{figure}

When we now include the magnet in the current path (Fig. 3), the differential conductance shows an even more complicated variation with carrier concentration. This is highlighted in panels (i) $-$ (v), in which $g_d$ either exhibits a local peak, or minimum, at zero bias. While differential-conductance measurements have previously been used to explore the influence of carrier heating in graphene \cite{Betz2012,Price2012}, those experiments reveal a slow variation of $g_d$ as a function of the applied bias. The behavior in Fig. 3 is very different, with $g_d$ exhibiting rich fine structure and, in many cases, an asymmetric response with regards to the DC-bias polarity. We attribute this response to the influence of the magnet, and to the capacity of the applied bias, over the narrow range considered here, to spectroscopically probe \cite{Somphonsane2020} the hybrid graphene system.

\begin{figure}[t]
\begin{center}
\includegraphics[width=0.99\columnwidth]{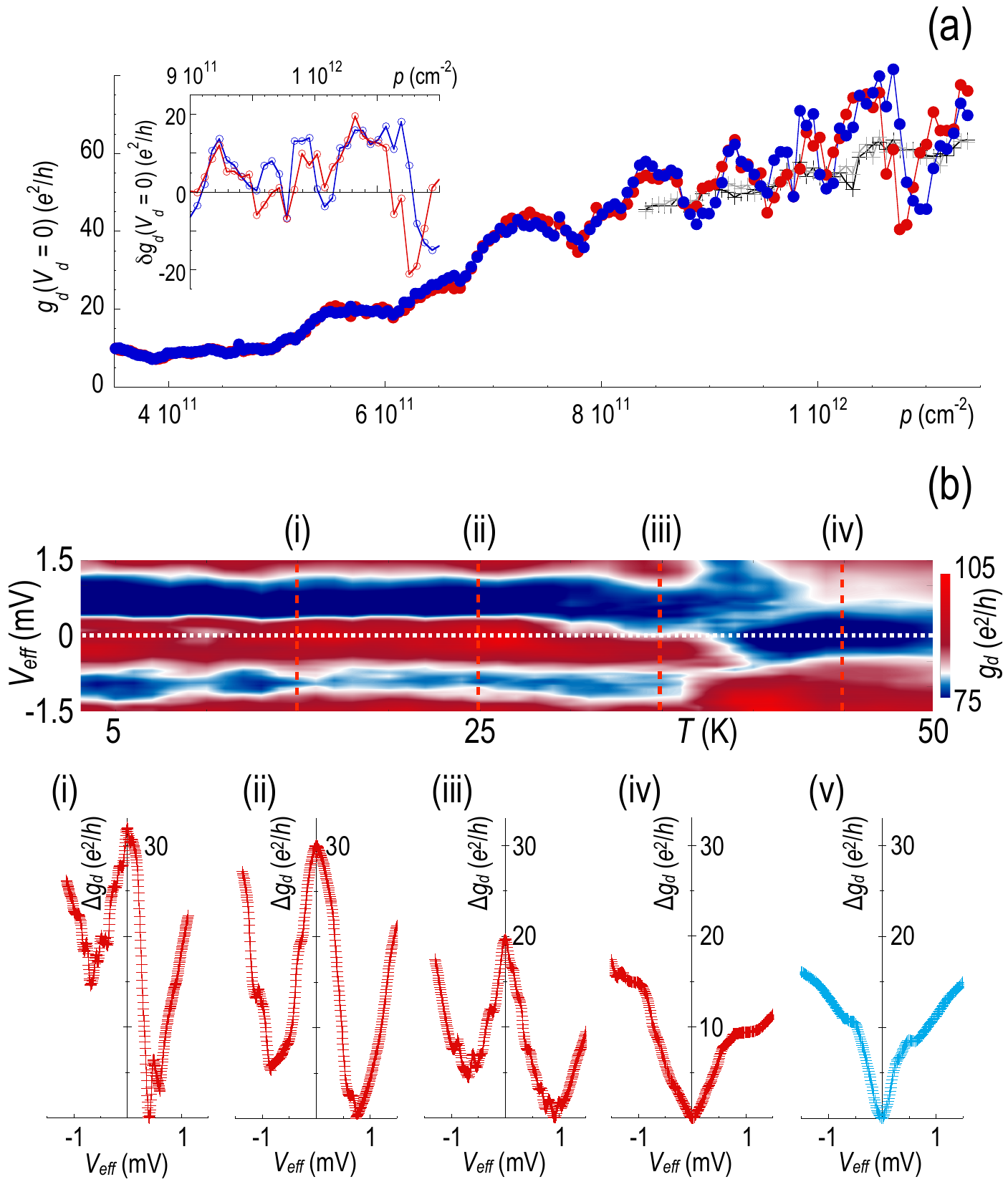}
\end{center}
\caption{
(Color online) (a) Variation of the zero-bias conductance ($g_d(V_d) = 0$) of I-D:1 as a function of hole concentration. Filled symbols (crosses) correspond to measurements performed with the magnetic element included in (excluded from) the current path. Red (black) data points correspond to measurements performed while sweeping the bias voltage up; down-sweep data are indicated in blue (gray). The inset plots the difference in the zero-bias conductance ($\delta g_d(V_{d}=0)$), from measurements performed with and without the magnet present. Red (blue) data were obtained by subtracting the sweep-up (sweep-down) measurements of the main panel. (b) The upper panel is a color contour showing the variation of $g_d$ as a function of the effective bias and temperature. Line plots (i) $-$ (iv) correspond to the temperatures (15-, 25-, 35- and 45-K, respectively) denoted by white dotted lines in the contour. Vertical scale is the same in all cases. Panel (v) shows, for comparison, a typical differential-conductance trace obtained (at 3 K) with the magnet excluded from the current path.
}
\label{Fig4}
\end{figure}


In the main panel of Fig. 4(a), we plot zero-bias conductance ($g_d(V_{d} = 0)$) as a function of $p$, with and without the magnet present. While reproducible fluctuations are present in both measurements, \textcolor{black}{with the magnet outside the current path (black and gray symbols) their RMS amplitude is around 2$e^2/h$ (see Section S4 the supplement); fluctuations of this size are common to experiments performed in a four-probe geometry, when transport over the region sampled by the voltage probes is phase coherent \cite{Benoit1987}.} Very different behavior occurs with the magnet present, however, in which configuration the fluctuations grow systematically with increasing carrier concentration. At the highest concentrations, the amplitude of fluctuation reaches as much as 40$e^2/h$, or roughly 50$\%$ of the background conductance, and representing a profound violation of the universal character of mesoscopic transport. An alternative means of demonstrating the nonuniversality is presented in the inset to Fig. 4(a), where we plot the difference in the zero-bias conductance ($\delta g_d(V_{d}=0) \equiv g_d(0,\mathcal{I} \textnormal{ present}) - g_d(0,\mathcal{I} \textnormal{ absent})$), determined from the two measurements in the main panel.

The temperature ($T$) dependence of the \textcolor{black}{signature observed in the detection-interaction geometry is demonstrated} in Fig. 4(b), the color contour of which shows the variation of differential conductance with temperature and bias for I-D:1. The line plots labeled (i) $-$ (iv) below the contour show measurements at temperatures identified in that panel. According to these data, \textcolor{black}{the differential conductance exhibits a zero-bias peak that} is largely unchanged up to $\sim$15 K, a value comparable to the width (1 $-$ 2  mV) of this feature. With further increase of temperature, the peak begins to weaken (see panel (iii)) and ultimately washes out around 40 K. At 45 K (panel (iv)), the peak is replaced by a zero-bias minimum, \textcolor{black}{similar to the localization-related feature \cite{Somphonsane2020} (panel (v)) that occurs with the magnet outside the circuit}. \textcolor{black}{This dip washes out with further increase of temperature beyond 50 K, consistent with our prior observations \cite{Somphonsane2017}}.

\textcolor{black}{Before proposing an explanation for our experiment, we first exclude some possibilities. One possibility is that fringing fields, emanating from the magnet into the detection region \cite{Pietrobon}, are responsible for our observations. We discount this, however, noting that the field strength a few microns away from the magnet should be weak \cite{Bae}. Additionally, if such fields were responsible for our observations, they should influence measurements in the same manner, with or without the magnet in the measurement circuit. As a second possibility, a zero-bias anomaly, reminiscent of that seen here, is a common signature of the Kondo effect \cite{Kondo1,Kondo2,Kondo3}. Our system involves the interaction of a conducting sea of carriers with an \emph{ordered ferromagnet}, however, unlike the scattering from a single spin (or an ensemble of isolated spins) involved in the Kondo effect \cite{Kondo1,Kondo2,Kondo3}.}

\textcolor{black}{The main finding of our study is that, in experiments performed in the interaction-detection geometry, the conductance of the detection region contains both local and remote components. The remote signatures include peak-like structures in the differential conductance, at or around zero bias, and giant fluctuations as a function of carrier density, all of which definitively arise from the proximity of graphene to the ferromagnet in the interaction region. The mesoscopic nature of the remote features is testified to by their stochastic evolution with gate voltage, pointing to a connection to phase-coherent transport. Indeed, the remote signatures persist over a similar range of temperature to the normal weak-localization effect \cite{Somphonsane2017,Somphonsane2020}, itself a signature of phase coherence in carrier diffusion.}

\textcolor{black}{In an idealized graphene/cobalt system, we have seen theoretically how the Dirac bands of graphene are modified by both exchange coupling and spin-orbit interaction (see Fig. 1(c) and Section S6 of the supplement). 
While these model calculations do not address all details of our experiments (including the relative orientation of the graphene and cobalt crystal structures, and the influence of defects and the number of graphene layers), they nonetheless suggest the presence of an externally induced spin-orbit coupling (SOC) and spin polarization in the graphene.}
A well known consequence of SOC in quantum diffusion is a conversion of weak localization into anti-localization \cite{Bergmann1984,McCann2006}. In many respects this is reminiscent of our experiments, in which adding the magnet into the measurement circuit generates conductance peak(s) at/around zero bias. 
Noting this, we suggest that the peak(s) may result from anti-localization of carriers, arising from the SOC induced in the graphene by the ferromagnet.
In mesoscopic systems in general, it is well known that (anti-) localization can be non-self-averaging, and evolve stochastically as a function of carrier density \cite{Chang1994,Chan1995,Huibers1998}; by changing phase interference among the different partial waves responsible for the localization, this gives rise to random, yet deterministic, changes in the localization signal. Collecting these ideas, we suggest that the large, non-universal, conductance fluctuations in Fig. 4(a) actually arise from the mixing of a remotely-generated anti-localization signature with the local conductance. Unlike the local fluctuations, the remote contribution to the conductance due to anti-localization need not be bound by any universal value, and it is this aspect that we suggest leads to the large fluctuations seen in experiment \cite{Endnote}. While quantitative theory of this effect is currently lacking, this does not alter the fact that what we achieve in our experiment is remote detection of the artificial \emph{magnetic texture} (SOC and spin polarization), induced in graphene by the ferromagnet. 

\textcolor{black}{In Section S5 of the supplement, we show similar phenomena to those discussed here in an additional device (I-D:2). With the magnetic element included in the measurement, this device also exhibits a differential conductance peak(s) near zero bias, with a lineshape that again evolves in a complicated manner as the gate voltage is varied. In contrast to the device studied in the main paper, however, the features observed in I-D:2 are present on both the electron and hole sides of the Dirac point. Such sample-to-sample variability is typical of phenomena involving the quantum interference of carriers \cite{Lee1987}.}

\textcolor{black}{In conclusion, we have demonstrated a pronounced breakdown of the universal character of quantum transport in a graphene-based interaction-detection geometry. This setup allows us to induce an artificial magnetic texture in the graphene, and demonstrates how local considerations can be overwhelmed by remote signatures in phase-coherent conductors.}

Work at Buffalo was supported by the U.S. Department of Energy, Office of Basic Energy Sciences, Division of Materials Sciences and Engineering (DE-FG02-04ER46180). Theory was supported by the National Science Foundation (NSF-ECCS 1740136), and by nCORE, a wholly owned subsidiary of the Semiconductor Research Corporation (SRC), through the Center on Antiferromagnetic Magneto-electric Memory and Logic (tasks 2760.001 and 2760.002). JF acknowledges support from the Swedish Vetenskapsr\aa det.


\begin{thebibliography}{99}
\bibitem{Reproducibility} \textcolor{black}{Fast fluctuations of the magnetic state of the cobalt element, occurring much faster than our measurements, cannot be excluded here, but could be resolved in studies performed in the presence of an external magnetic field.}
\end{thebibliography}
\end{document}